\documentclass[
preprintnumbers,
prl,
a4paper,
twocolumn,
showpacs,
amsmath,
amssymb,
byrevtex
]{revtex4}
\usepackage{graphicx}
\begin{document}
%%%%%%%%%%%%%%%%%%%%%%%%%%%%%%%%%%%%%%%%%%%%%%%
\title{High-energy head-on collisions of particles and hoop conjecture}
\author{Hirotaka Yoshino}
\email{hyoshino@allegro.phys.nagoya-u.ac.jp}
\author{ Yasusada Nambu}
\email{nambu@allegro.phys.nagoya-u.ac.jp}
\affiliation{Department of Physics, Graduate School of Science, Nagoya
University, Chikusa, Nagoya 464-8602, Japan}
%%%%%%%%%%%%%%%%%%%%%%%
\preprint{DPNU 02-09}
\date{April 19, 2002}
%%%%%%%%%%%%%%%%%%%%%%%
\begin{abstract}
We investigate the apparent horizon formation for high-energy head-on
collisions of particles in multi-dimensional spacetime. The apparent
horizons formed before the instance of particle collision are obtained
analytically. Using these solutions, we discuss the feature of the apparent
horizon formation in the multi-dimensional spacetime from the
viewpoint of the hoop conjecture. 
\end{abstract}
%%%%%%%%%%%%%%%%%%%%%%%
\maketitle
%%%%%%%%%%%
{\it Introduction.} --- %
%%%%%%%%%%%
Brane world scenario has been discussed by many authors recently. The
scenario suggests that the Planck energy can be as low as
$O(\text{TeV})$ scale~\cite{TSS}. If the Planck energy is TeV scale, it is
possible to create black holes using accelerators, such as
LHC~\cite{BHUA}. Further, the collisions of cosmic rays with
our atmosphere have energy reach beyond that of LHC and their observation 
will find the existence of the large extra-dimension or
will place improved bounds on the fundamental Planck scale.
Hence we would like to better understand the process of the black hole
formation via particle collisions. For this purpose, we investigate
the formation of the apparent horizon for the system of the
head-on collisions of high-energy particles.   

To simplify the analysis, we follows the method adopted by Eardley and
Giddings~\cite{EG02}.  First, the tension of the brane which is 
expected to be the Planck scale can be negligible if the center of
mass energy is substantially larger than the Planck scale. Second, the
geometry of the extra dimensions plays no essential role if the
geometrical scales of the extra dimensions are large compared to 
the horizon radius for the center of mass energy. Thus we consider
the head-on collisions in $D$-dimensional Einstein gravity.
The metric with a high-energy point particle is obtained by infinitely
boosting the Schwarzschild black hole metric with the fixed total energy
$\mu$. The resulting system becomes a massless point particle
accompanied by a plane-fronted gravitational shock wave which is the
Lorentz-contracted longitudinal gravitational field of the
particle. Combining two shock waves, we can set up the 
high energy collision. This system was originally developed by 
D'Eath and Payne~\cite{DEP92}. The black hole formation
with an impact parameter for $D=4$ was investigated by Eardley and
Giddings~\cite{EG02}, and they showed that the apparent horizon which
encloses two particles exists at the instance of collision
for sufficiently small impact parameter. 

We examine the head-on collisions using the different slicing of the spacetime:
 we expect that the apparent horizon forms before the collision
of two particles. We  construct the solutions of the
apparent horizons analytically and discuss how the dimension $D$
affects the formation of the horizon from the viewpoint of the hoop
conjecture~\cite{Th72}.  

%%%%%%%%%%%%%%%%%%%%%%%%%%%%%%
{\it High-energy particle collisions at the speed of light.} --- %
%%%%%%%%%%%%%%%%%%%%%%%%%%%%%%
The gravitational solution for the each incoming particles can be
found by boosting the rest-frame $D$-dimensional Schwarzschild
solution, 
\begin{align}
&ds^2=-\left(1-\frac{16\pi G_DM}{(D-2)\Omega_{D-2}}\frac{1}{r^{D-3}}\right)dt^2
\nonumber\\
& +\left(1-\frac{16\pi
    G_DM}{(D-2)\Omega_{D-2}}\frac{1}{r^{D-3}}\right)^{-1}dr^2
+r^2d\Omega_{D-2}^2, 
\end{align}
where $d\Omega_{D-2}^2$ and $\Omega_{D-2}$ are the line element and
volume of the unit $D-2$-sphere and $G_D$ is the $D$-dimensional 
gravitational constant. Aichelburg-Sexl solution~\cite{AS71} is found
by taking the limit of large boost and small mass with the fixed total
energy $\mu$. The resulting metric represents a massless particle
moving in the $+z$ direction with the speed of light: 
\begin{equation}
ds^2=-d\bar{u} d\bar{v}+\sum_{i=1}^{D-2}d\bar{x}_i^2+\Phi(\bar{x}_i)
\delta(\bar{u})d\bar{u}^2,
\label{eq:delta}
\end{equation}
where $\bar{u}=\bar{t}-\bar{z}$ and $\bar{v}=\bar{t}+\bar{z}$.
$\Phi$ depends only on the transverse radius 
$\bar{\rho}=\sqrt{\bar{x}_i\bar{x}^i}$ and takes the form
\begin{align}
& \Phi=-8G_4\mu\log\bar{\rho},\quad\text{for}~D=4, \\
& \Phi=\frac{16\pi\mu
  G_D}{\Omega_{D-3}(D-4)}\frac{1}{\bar{\rho}^{D-4}},\quad\text{for}~D>4.
\end{align}
A delta function appeared in \eqref{eq:delta} shows that two
coordinate systems are discontinuously connected on $\bar{u}=0$. The continuous
coordinate system can be introduced by 
\begin{align}
& \bar{u}=u, \notag\\
& \bar{v}=v+\Phi\theta(u)
+\frac{u}{4}\theta(u)\left(\nabla_i\Phi\nabla^i\Phi\right),\\
& \bar{x}_i=x_i+\frac{u}{2}\nabla_i\Phi(x_i)\theta(u), \notag
\end{align}
where $\theta$ is the Heaviside step function and $\nabla_i$ is the
$(D-2)$-dimensional flat-space derivative. 
We can superpose the two solutions to obtain the exact geometry outside
the future light cone of the collision of the shocks:
\begin{align}
&ds^2=-dudv\nonumber\\
&+\left(H^{(1)}_{ik}H^{(1)}_{jk}+H^{(2)}_{ik}H^{(2)}_{jk}
-\delta_{ij}\right)dx^idx^j,
\end{align}
where
\begin{align}
&H^{(1)}_{ij}=\delta_{ij}+\frac{u}{2}\theta(u)\nabla_i\nabla_j\Phi^{(1)}(\boldsymbol{x}),
\notag \\
&H^{(2)}_{ij}=\delta_{ij}+\frac{v}{2}\theta(v)\nabla_i\nabla_j\Phi^{(2)}(\boldsymbol{x}).
\end{align}
Here $\boldsymbol{x}\equiv(x^i)$ is the point in flat $D-2$-space
that is transverse to the direction of particle motion.

The apparent horizon is defined as a closed spacelike $D-2$-surface 
on which the outer null geodesic congruence have zero convergence.
It was shown that the apparent horizon exists in the union of the two
shock waves, $u=0>v$ and $v=0>u$~\cite{EG02}.
This apparent horizon consists of two flat discs with  radii
\begin{equation}
\rho_0\equiv \left(\frac{8\pi\mu G_D}{\Omega_{D-3}}\right)^{1/(D-3)},
\end{equation}
and $\rho_0$ gives a characteristic scale for each dimension $D$.

%%%%%%%%%%%%%%%%%%%%%%
{\it Time slicing and apparent horizons.} ---%
%%%%%%%%%%%%%%%%%%%%%%
To treat the collision of particles as time evolutional process,
we consider the following slice of spacetime:
\begin{align}
&\text{region I}: t=z, t\le T, \notag \\
&\text{region II}: z=-t, t\le T,\\
&\text{region III}: t=T, -T\le z\le T, \notag
\end{align}
%~\cite{Th72}
where $T\le 0$ and particles collide at $T=0$.
In order to find apparent horizon on the above slice, we first prepare surfaces
with zero expansion in region I and III,  then connect them
smoothly by requiring that the null normal coincides at the junction
of region I and III, $t=z=T$. In region I, the surface which have zero
expansion is given by 
\begin{equation}
v=-\Phi+\text{const.},
\end{equation}
and its null normal $k_1^a$ is
\begin{align}
& k_1^u=\left({\rho_0}/{\rho}\right)^{-(D-3)}, \notag \\
& k_1^v=\left({\rho_0}/{\rho}\right)^{D-3},\\
& k_1^\rho=1. \notag
\end{align}
In region III, the surface which have zero expansion is given by
\begin{equation}
az=\pm f(a\rho),
\end{equation}
where $a$ is a constant of integration determined by the matching condition
at the junction. For $D=4$, the function $f(x)$ is given by
\begin{equation}
f(x)=\cosh^{-1} x,
\end{equation}
and for $D>4$,
\begin{align}
f(x)&=-\frac{x^{-D+4}}{D-4}
 {}_2F_1\left(\frac{1}{2},\frac{D-4}{2(D-3)},
  \frac{3D-10}{2(D-3)},x^{2(3-D)}\right)   \notag\\
&\qquad\qquad\qquad\qquad
-\sqrt{\pi}\frac{\varGamma\left(\frac{D-4}{2(D-3)}\right)}{\varGamma\left(\frac{1}{2(3-D)}\right)},
\end{align}
where ${}_2F_1$ is the Gauss' hyper-geometric function. The null
normal $k_3^a$ of the surface is given by 
%%%
\begin{align}
& k_3^u=(a\rho)^{D-3}-\sqrt{(a\rho)^{2(D-3)}-1},\notag\\
& k_3^v=(a\rho)^{D-3}+\sqrt{(a\rho)^{2(D-3)}-1},\\
& k_3^\rho=1. \notag
\end{align}
%%%
Matching these surfaces and null normals at the junction $t=z=T$, we have
\begin{align}
f(a\rho_b)&=-aT,\\
\left({\rho_0}/{\rho_b}\right)^{D-3}&
 =(a\rho_b)^{D-3}+\sqrt{(a\rho_b)^{2(D-3)}-1}.
\end{align}
where $\rho_b$ is the radius of the surface at the junction. From
this, the relation between $T$ and $\rho_b$ can be given parametrically as
\begin{align}
&\frac{T}{\rho_0}=-\xi
f\left(\frac{1}{\xi}(2\xi^{3-D}-1)^{1/2(3-D)}\right)
 , \\
&\frac{\rho_b}{\rho_0}=\left(2\xi^{3-D}-1\right)^{1/2(3-D)}.
\end{align}
where $0\le\xi\le 1$.
%%%% fig1%%%%%
\begin{figure}[b]
\centering
\includegraphics[width=1.0\linewidth,clip]{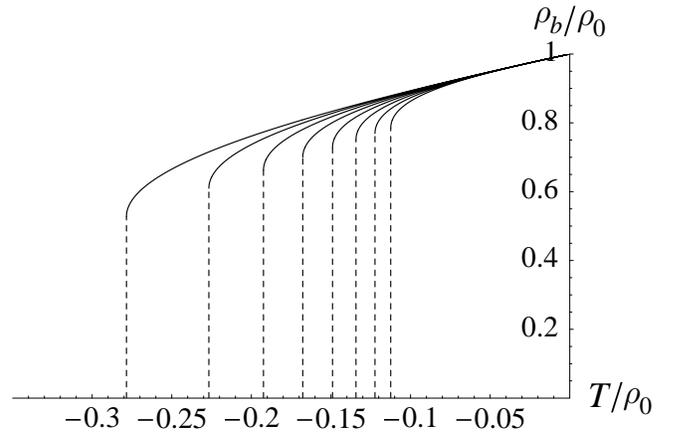}
\caption{ The relation between $T$ and the horizon radius $\rho_b$ for
  $D=4,\cdots, 11$. The intersection of the dotted line and $T/\rho_0$-axis is 
the time $T_{c}$ when the apparent horizon appears.
The value of $|T_c/\rho_0|$ decreases as $D$ increases.}
\label{fig:fig1}
\end{figure}
%%%%%%%%%%%
FIG. 1 shows the relation between $T$ and $\rho_b$ for each $D$. We denote the
time when the apparent horizon appears as $T=T_{c}$. The value of
$|T_c/\rho_0|$ becomes small as $D$ increases.  For large $D$, we have
%%%
\begin{equation}
 \rho_b/\rho_0\approx D^{-1/2D},\quad T_c/\rho_0\approx -1/D
\end{equation}
%%%
at $T=T_c$. 
The intersection of  the $z=\text{const.}$ plane and the surface in region III
is a $D-3$-dimensional sphere, of which expansion is positive and 
proportional to $D-3$.
Thus the surface has negative expansion on
$(\rho/\rho_0,z/\rho_0)$-plane and 
its curvature on this plane increases with the increase of space-time dimension $D$.
This leads to the decrease in the distance of two particles at the  horizon formation.
 The shape of apparent horizons for $D=4$ and $D=5$ are shown in FIG. 2 and FIG. 3. 

%%%%%% fig2 %%%%%
\begin{figure}[t]
\centering
\includegraphics[width=1.0\linewidth,clip]{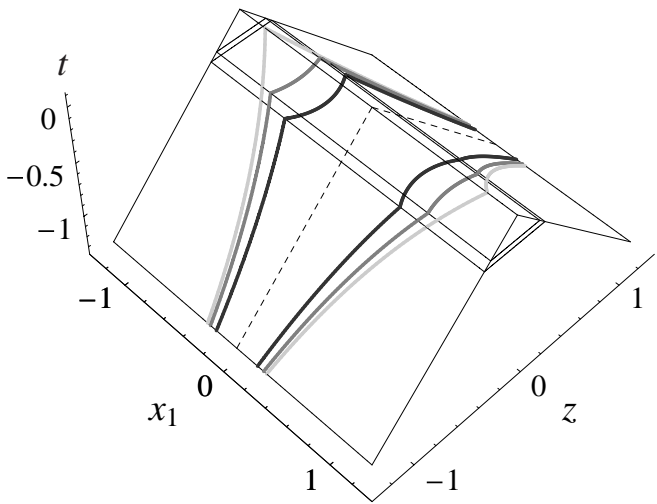}
\caption{ The apparent horizon for $D=4$ at $T/\rho_0=-0.278,-0.225,0$.
The dark line is the horizon at $T=T_c=-0.278\rho_0$, and light line is the
horizon at $T=0$. The unit of the axis is $\rho_0$. }
\label{fig:fig2}
\end{figure}
%%%%%%%%%%%%%%

%%%%%%%%%%%%%
{\it Hoop conjecture.} ---%
%%%%%%%%%%%%%
Now we examine the difference of the horizon formation for  various
spacetime dimension using the hoop conjecture. The hoop conjecture
gives the criterion of black hole formation in the 4-dimensional general
relativity~\cite{Th72}. It states that an apparent horizon forms when and only when
 the mass $M$ of the system gets compacted into a region of which circumference $C$ satisfies
\begin{equation}
H_4\equiv C/4\pi G_4 M\lesssim 1.
\end{equation} 
 As $4\pi G_4
M$ is the circumference of the 4-dimensional Schwarzschild horizon, 
we can  expect that the criterion of black hole formation in the  
$D$-dimensional Einstein gravity is given by  
\begin{equation}
H_D\equiv C/2\pi r_{\text{h}}(M)\lesssim 1,
\end{equation} 
where $r_{\text{h}}(M)$ is the Schwarzschild radius of $D$-dimensional
spacetime. This criterion was implicitly used to
estimate the total cross section for black hole production 
via non-head-on collisions~\cite{BHUA}.

To calculate the ratio $H_D$ and $H_4$, we must specify the definition of
the mass of the system. In this paper, we use total energy $E=2\mu$ as
the mass of the system. The circumference $C$ is defined as minimum
length which encloses two particles. We take the loop as shown in
FIG. 4 and calculate $C$  by taking the limit $c\rightarrow 0$.
$C$ reduces to $4|T|$ which is the twice the distance of two particles.
The value of $H_D$ at $T=T_c$ is shown in TABLE I. As $D$ increases,
the value of $H_D$ decreases and the mass $M$ must be compacted into 
the region with smaller circumference $C$ than $2\pi r_{\text{h}}$ 
to produce a black hole. 
This reflects the decrease in $|T_c|/\rho_0$ with increase in $D$.

%%%%% fig3 %%%%
\begin{figure}[t]
\centering
\includegraphics[width=1.0\linewidth,clip]{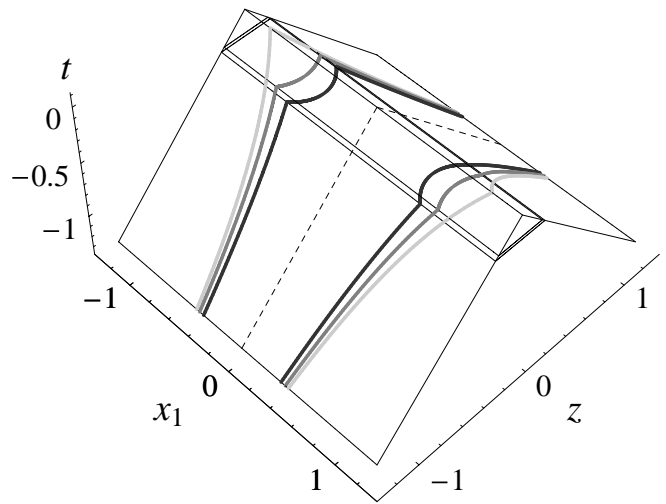}
\caption{ The apparent horizon for $D=5$ at $T/\rho_0=-0.227,-0.2,0 $.
The dark line is the horizon at $T=T_c=-0.227\rho_0$, and light line is the 
horizon at $T=0$. The unit of the axis is $\rho_0$.}
\label{fig:fig3}
\end{figure}
%%%%%%%%%%%%

%%%%% fig4 %%%%
\begin{figure}[b]
\centering
\includegraphics[width=1.0\linewidth,clip]{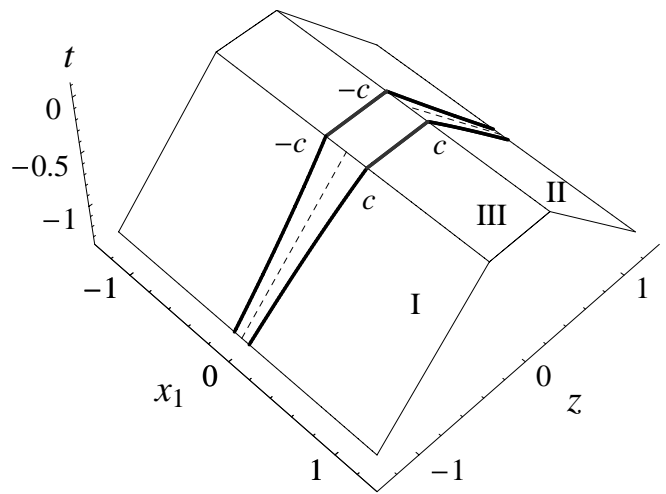}
\caption{ The closed loop to calculate the circumference.
We calculate $C$ by taking $c\rightarrow 0$.}
\label{fig:fig4}
\end{figure}
%%%%%%%%%%%%

The value of $H_4$ at $T=T_c$ is also shown in TABLE I.
This result can be written as 
\begin{equation}
H_4=F(D)\frac{(G_DE)^{1/(D-3)}}{G_4E},
\label{eq:hoop4}
\end{equation}
where $F(D)=0.03\sim 0.2$. The $D$-dimensional gravitational constant
is related to the Planck energy as
\begin{equation}
M_p^{D-2}=\frac{(2\pi)^{D-4}}{4\pi G_D}.
\end{equation}
Using this formula, Eq.\eqref{eq:hoop4} becomes
\begin{align}
& H_4=F(D)\left(\frac{M_4}{M_p}\right)^2
\left(\frac{8\pi^2M_p}{E}\right)^{\frac{D-4}{D-3}}.
\end{align}
If the Planck energy is TeV scale, $M_4/M_p$ is $\sim 10^{16}$ and
$H_4$ becomes  $\sim 10^{32}$. Thus the mass does not need to be
compacted into a small region of which circumference is 
 $C\lesssim 4\pi G_4M$ to produce a black hole.

%%%%%%%%%%%%%%%%%%%%%%%%%%%%%%%%%
\begin{table}
\caption{ The value of $H_4$ and $H_D$ at $T=T_c$ for $D=4\sim 11$.}
\begin{ruledtabular}
\begin{tabular}{rcc}
$D$ & $H_D$ & $H_4$\\
\hline
$4$ & $ 0.1773 $   &   $ 0.1773 $ \\
$5$ & $0.1567$ & $ 0.0722 \left[(G_5E)^{1/2}/(G_4E)\right]$ \\
$6$ & $0.1348$ & $ 0.0527 \left[(G_6E)^{1/3}/(G_4E)\right]$ \\
$7$ & $0.1176$ & $ 0.0444 \left[(G_7E)^{1/4}/(G_4E)\right]$ \\
$8$ & $0.1042$ & $ 0.0396 \left[(G_8E)^{1/5}/(G_4E)\right]$ \\
$9$ & $0.0936$ & $ 0.0364 \left[(G_9E)^{1/6}/(G_4E)\right]$ \\
$10$  & $0.0849$ & $ 0.0340 \left[(G_{10}E)^{1/7}/(G_4E)\right]$ \\ 
$11$  & $0.0777$ & $ 0.0321 \left[(G_{11}E)^{1/8}/(G_4E)\right]$\\
\end{tabular}
\end{ruledtabular}
\end{table}
%%%%%%%%%%%%%%%%%%%%%%%%%%%%%%%%%

%%%%%%%%%%%%%%%%%%
{\it Summary and discussion.} ---  %
%%%%%%%%%%%%%%%%%%
We have  investigated the temporal evolution of the apparent horizon for
high energy particle collisions. The apparent horizon which encloses
the two particles appears at $T=T_c$. Its radius increases in time and
reaches $\rho_0$ at $T=0$.  We calculated $H_D$ and found that
$H_D$ decreases as $D$ increases. This means that if we increase the space-time
dimension, the size of the hoop which enclose the
system should be much smaller than $2\pi r_{\text{h}}$. 
Therefore, the formation of the apparent horizon becomes more
difficult for larger $D$. On the other
hand,  $H_4=H_D\cdot r_{\text{h}}/2G_4M$ gives
 a large value  $\sim 10^{32}$ regardless of the decrease in $H_D$.
This is because the horizon radius $r_{\text{h}}$ becomes 
far larger than $2G_4 M$. As the horizon radius corresponds to 
the length scale which enclose the system, this leads to the conclusion that
 a black hole is easily formed in the TeV scale scenario.

Finally we discuss the validity of  the hoop conjecture. Obviously,
$H_4$ does not give the picture 
of the hoop conjecture because its value at the horizon formation is
far larger than unity. The ratio $H_D$ also does not  give the picture of the
hoop conjecture because its value at the horizon formation is much
smaller than unity. However, we used the rough estimated values of the
circumference $C$ and the mass $M$ to evaluate $H_D$ and $H_4$. 
The energy of shock wave with a high-energy particle is
distributed in the transverse direction of the motion, and our estimation of
the circumference $C$ is too small because the region surrounded by
this circumference does not enclose much of the gravitational energy.  
In our previous paper~\cite{YNT01}, we stated that $H_4$
with Hawking's quasi-local mass $M_{\text{H}}(\text{S})$~\cite{H68} 
becomes a  better parameter to judge the horizon formation for the system
with motions. We must calculate 
$H^{(\text{H})}_4(\text{S})=C(\text{S})/4\pi G_4 M_{\text{H}}(\text{S})$ 
for all surfaces S and then take the minimum value of them.
Even if the Hawking mass in multi-dimensional space-time has not been
calculated in this paper, we expect that $H^{(\text{H})}_D\lesssim 1$ becomes a
condition for the horizon formation. The value $H_D$ would decrease as $D$
increases even if we use the quasi-local mass because $H_D$ should reflect
the decrease in $|T_c|/\rho_0$. 

Although we can regard $C/2\pi r_{\text{h}}(M)\lesssim 1$ as  the
condition for the horizon formation in $D$-dimensional gravity, it
does not give a unique condition. The topology of apparent horizon is
not restricted to be $S^{D-2}$ surface in a multi-dimensional
space-time. Emparan and Reall derived  the solution of rotating black
ring in $D=5$~\cite{ER02}. For apparent horizon which does not  have
$S^{D-2}$ topology, the criterion for its formation may take another
form. Our criterion $C/2\pi r_h(M)\lesssim 1$ is  applicable only to
the horizon with $S^{D-2}$ topology. 

The authors would like to thank Akira Tomimatsu and Masaru Shibata for
helpful discussions. 

%%%%%%%%%%%%%%%%%%%%%%%%%%%%%%%%%%%%%%%%%%%%

\end{document}